\documentclass[twocolumn]{aastex63}
\usepackage[dvipdfmx]{}
\usepackage{pslatex} 

\newcommand{\twelveco}{\mbox{$^{12}$CO}} 
\newcommand{\thirteenco}{\mbox{$^{13}$CO}} 
\newcommand{\ceighteeno}{\mbox{C$^{18}$O}} 
\newcommand{\twelvecol}{\mbox{$^{12}$CO($J$=1--0)}} 
\newcommand{\thirteencol}{\mbox{$^{13}$CO($J$=1--0)}}

\newcommand{\ceighteenol}{\mbox{C$^{18}$O($J$=1--0)}} 
\newcommand{\ceighteenoh}{\mbox{C$^{18}$O($J$=2--1)}} 
\newcommand{\ceighteenohh}{\mbox{C$^{18}$O($J$=3--2)}}

\newcommand{\sioh}{\mbox{SiO($J$=5--4)}}

\newcommand {\msun}{\mbox{M$_\odot$}}
\newcommand {\lsun}{\mbox{L$_\odot$}}
\newcommand {\kms}{\mbox{km~s$^{-1}$}}

\newcommand {\vlsr}{\mbox{$V_{LSR}$}}
\newcommand {\nhtwo}{\mbox{$N_\mathrm{H_2}$}}
\newcommand {\nco}{\mbox{$N_\mathrm{C^{18}O}$}}
\newcommand {\tmb}{\mbox{$T_\mathrm{mb}$}}

\newcommand {\vcen}{\mbox{$V_\mathrm{center}$}}
\newcommand {\tex}{\mbox{$T_\mathrm{ex}$}}
\newcommand {\tbg}{\mbox{$T_\mathrm{bg}$}}
\newcommand {\tobs}{\mbox{$T_\mathrm{obs}$}}
\newcommand {\kb}{\mbox{$k_\mathrm{B}$}}
\newcommand {\mum}{\mbox{$\mu_\mathrm{m}$}}
\newcommand {\mpro}{\mbox{$m_\mathrm{p}$}}
\newcommand {\dv}{\mbox{$dV$}}
\newcommand {\hone}{\mbox{H{\sc i}}}
\newcommand {\htwo}{\mbox{H{\sc ii}}}
\newcommand {\rco}{\mbox{R$_{2-1/1-0}^{18}$}}

\accepted{ApJ}
\shorttitle{CCC in Sgr\,B}
\shortauthors{Enokiya et al.}

\begin{document}

\title{A Multiwavelength Study of the Sgr\,B Region: Contiguous Cloud-Cloud Collisions Triggering Widespread Star Formation Events?}

\correspondingauthor{Rei Enokiya}
\email{enokiya@a.phys.nagoya-u.ac.jp}

\author[0000-0003-2735-3239]{Rei Enokiya}
\affiliation{Department of Physics, Nagoya University, Chikusa-ku, Nagoya, Aichi 464-8601, Japan}
\affiliation{Department of Physics, Faculty of Science and Technology, Keio University, 3-14-1 Hiyoshi, Kohoku-ku, Yokohama, Kanagawa 223-8522, Japan}
\author{Yasuo Fukui}
\affiliation{Department of Physics, Nagoya University, Chikusa-ku, Nagoya, Aichi 464-8601, Japan}

\begin{abstract}
The Sgr\,B region, including Sgr\,B1 and Sgr\,B2, is one of the most active star-forming regions in the Galaxy. \citet{has94} originally proposed that Sgr\,B2 was formed by a cloud-cloud collision (CCC) between two clouds with velocities of $\sim$45 $\kms$ and $\sim$75 $\kms$. However, some recent observational studies conflict with this scenario.
We have re-analyzed this region, by using recent, fully sampled, dense-gas data and by employing a recently developed CCC identification methodology, with which we have successfully identified more than 50 CCCs and compared them at various wavelengths.
We found two velocity components that are widely spread across this region and that show clear signatures of a CCC, each with a mass of $\sim$10$^6$ $\msun$.
Based on these observational results, we suggest an alternative scenario, in which contiguous collisions between two velocity features with a relative velocity of $\sim$20 $\kms$ created both Sgr\,B1 and Sgr\,B2.
The physical parameters, such as the column density and the relative velocity of the colliding clouds, satisfy a relation that has been found to apply to the most massive Galactic CCCs, meaning that the triggering of high-mass star formation in the Galaxy and starbursts in external galaxies can be understood as being due to the same physical CCC process.
\end{abstract}

\section{Introduction} \label{sec:intro}
\subsection{The Center of the Galaxy}
Despite its small volume,  the inner 1 kpc of the Galaxy contains several to 10$\%$ of molecular gas, and thus it is a prominent mass reservoir \citep{mor96}.
As expected from the large amount of gas, the strength of the frozen-in magnetic fields is very high in the Galactic Center (GC).
It has been estimated to be at least 30 $\mu$G \citep{cro10} and this leads to very large Alfv\'en speeds of a few tens to a hundred $\kms$.
Thus, the nonnegligible field can accelerate gas, and this makes the situation more complicated in the GC \citep{fuk06}.

Molecular-line observations have revealed that the molecular gas in the GC exhibits a very large velocity width and a parallelogram shape in the $l$--$v$ diagram, which cannot be explained by Galactic rotation \citep[e.g.,][]{oka98, tsu99}.
The acceleration mechanism of the gas is still unclear, but so far two possible accelerators have been suggested.

One is the bar potential, originally proposed by \citet{bin91}.
They suggested that the framework of the parallelogram in the $l$--$v$ diagram is explained by the innermost orbits of the so-called $x_{1}$ orbits.
However, there has been no firm observational evidence of the small bar, which is required to fit the size of the parallelogram to the observed one.
Furthermore, many recent theoretical and observational works have revealed that the bar is much larger and rotates at a much slower pattern speed than assumed in \citet{bin91} \citep[see e.g.,][]{weg13,sor15,por17,san19}.

In accordance with those findings, \citet{fux99} performed simulations using a composite $N$-body and hydro code, and exhibited asymmetric and non-stationary gas flows that substantially reproduce the observed $l$--$v$ diagrams of $\hone$ and CO.
Recent hydrodynamic simulations by \citet{rod11} and \citet{sor18} have successfully explained the lateral sides of the parallelogram as dust lanes or gas that has been collisionally accelerated by dust lanes.
On the other hand, the top and bottom sides of the parallelogram are still unclear.
Although \citet{sor18} have interpreted them as brushed gas between the dust lanes and the central molecular zone (CMZ), the boundaries of the top and bottom sides in the simulations still remain obscured and are not so clear compared to the CO emissions seen in \citet{oka98}.

The other acceleration mechanism involves magnetic instabilities, such as the magnetorotational instability and Parker instability originally proposed by \citet{fuk06}.
Some magnetohydrodynamic (MHD) simulations have revealed that this mechanism can also accelerate gas and can reproduce a parallelogram-like distribution as a transient feature \citep[e.g.,][]{suz15}, although the shape is blobby and not clear.
This model also has observational difficulty.
Obtaining a direct detection of the magnetic field embedded in the dense molecular gas with a large line width is fairly difficult, so there have been few works that have tackle the situation through a new method, using far-infrared (FIR) polarimetry and velocity gradients of gas \citep{hu21}.

Similar to the acceleration mechanism, many models for the dynamical structure of the molecular condensation within the central 200 pc---hereafter, the CMZ  \citep{mor96}---have been proposed (e.g., an expanding shell---\citealt{sof95}; a twisted ring---\citealt{mol11}; and an open orbit---\citealt{kru15}).
Regardless of which model is ultimately proven to be correct, revealing the star formation mechanism in the dense molecular cloud in such an extreme magnetized, highly pressured environment is important for our understanding of the Galaxy evolution.

Although the central mass reservoir often acts as a star formation engine, star formation in the GC is inactive compared to the theoretical expectation, and possibly suppressed \citep[e.g.,][]{lon13}.
In order to unveil this star formation mystery, a multiwavelength study---including both dense- and diffuse-gas tracers---will be important, because the star formation timescale is shortened in such extreme condition by external pressures \citep{kau16}, and hence the dense gas no longer holds information regarding the triggering mechanism.
Thanks to the improvement in heterodyne receivers over the last two decades, plenty of molecular line surveys have been carried out \citep[e.g.,][]{jon12}.
However, most works focus only on the main streams of the CMZ, with dense-gas tracers, and do not include the diffuse gas traced by CO.
We therefore focus on the most active star forming region in the CMZ, Sgr\,B, and conduct a multiwavelength study with the latest archival data sets, which cover large area and density range, integrating them with knowledge obtained from previous studies.

\subsection{The Sgr\,B region}
The Sgr\,B region, consisting of Sgr\,B1 and B2, is one of the brightest radio sources, and is located a projected distance of $\sim$100 pc away from the GC.
The Sgr\,B region is also known to be a strong SiO source in the Galaxy \citep{mar97}.
\citet{Meh92} carried out observations of the radio-continuum and recombination lines toward Sgr\,B1 using the Very Large Array (VLA), and they found extended emission in Sgr\,B1, unlike Sgr\,B2.
They discussed the possibility that the emission may have been caused by coalesced, evolved $\htwo$ regions, and from the proximities of Sgr\,B1 and Sgr\,B2 in terms of their projected distances and revealed ionized gas velocities, they suggested that Sgr\,B1 and Sgr\,B2 are a unified structure.
\citet{Meh93} investigated OH and H$_{2}$O masers toward Sgr\,B1 using the VLA, and discovered an OH maser and ten H$_{2}$O maser spots.
These observations provide evidence for ongoing star formation in Sgr\,B1, although the majority of the radio-continuum emission---or, in other words, the star-forming activity---in this region is dominated by evolved $\htwo$ regions.
Although estimating a quantitative value for the star-forming activity in the GC region is very difficult, due to the heavy extinction at optical wavelengths, \citet{sim18} found that at least eight separate subregions were highly excited by high-mass stars.
From their analyses of the spectral energy distributions obtained from [O{\sc iii}] 52 and 88 $\mu$m data, taken with the Stratospheric Observatory for Infrared Astronomy, they concluded that these high-mass stars were equivalent to late O-type stars.
They also estimated the ages of those late O-type stars to be at least a few Myr.
From analyses of radio-continuum emissions obtained with the Wilkinson Microwave Anisotropy Probe, \citet{bar17} also estimated the total stellar mass embedded within Sgr\,B1 to be $\sim$8000 $\msun$.

Sgr\,B2 is young, and it is the most active star-forming region in the Galaxy.
It contains numerous O-type stars, $\htwo$ regions, ultracompact and compact $\htwo$ regions, H$_{2}$O masers \citep[e.g., ][]{yus04,mcg04}, and also a wealth of chemical species, such as organic molecules.
The IR luminosity is incredibly high, $\sim$10$^7$ $\lsun$ \citep{gol87}, and its FIR spectrum and [C{\sc ii}] 157.7 $\mu$m line are comparable to those in FIR-bright galaxies \citep{vas02}.
Almost all of the star-forming activity is concentrated within three regions, with diameters of $\sim$1 pc, called Sgr\,B2(N), Sgr\,B2(M), and Sgr\,B2(S), which are aligned from north to south \citep{ben84}.
Sgr\,B2(M) is a notable starburst region that contains 49 ultracompact $\htwo$ regions \citep{gau95}.
Recently, \citet{gin18} carried out 3 mm radio-continuum observations with ALMA, and discovered 271 compact sources, which may consist of hypercompact $\htwo$ regions and massive young stellar objects, within a 5 pc radius around Sgr\,B2(M). 

Sgr\,B2 is associated with a giant molecular cloud containing 6 $\times$ 10$^{6}$ $\msun$ \citep{lis89}.
\citet{sof90} discovered giant molecular complexes associated with Sgr\,B1 and B2, namely, the Sgr\,B1 molecular shell and the Sgr\,B2 molecular complex.
He suggested that star formation in Sgr\,B2 was triggered by a compression induced by the expansion of the Sgr\,B1 molecular shell, driven by feedback from Sgr\,B1. 
On the other hand, \citet{has94} proposed a cloud-cloud collision (hereafter, CCC) scenario for the Sgr\,B2 region, based on their $\thirteencol$ observations with the Nobeyama 45 m telescope.
They discovered three clouds in the direction of this region, which they named the Shell, at $\vlsr$ = 20--40 $\kms$; the Hole, at $\vlsr$ = 40--50 $\kms$; and the Clump, at $\vlsr$ = 70--80 $\kms$ (see also Figure~\ref{lbch}).
They interpreted the Clump, Hole, and Shell as clouds that have been compressed, hollowed, and shocked by the CCC, respectively, and they concluded that massive star formation in the Sgr\,B2 complex was triggered by the collision.
\citet{sat00} and \citet{has08} lent further support to this model, by comparing the spatial and velocity distributions of their $\ceighteenol$, $\ceighteenohh$, CS($J$=1--0), and CS($J$=7--6) data sets with those of the masers.
However, these models are based on under-sampled datasets, so the effective spatial resolution is low. 
Also, this scenario conflicts with recent observational results at other wavelengths.
By analyzing HNCO data, \citet{hen16} found that there is only one cloud toward Sgr\,B2, and they suggested that the cloud identification by \citet{has94} was suspicious.
Based on a dynamical model of the CMZ, \citet{kru19} suggested that the complex kinematic structure of Sgr\,B2 was naturally explained by the superimposition of fragmentation and orbital motion of a cloud in the line of sight (LOS).

Although Sgr\,B2 is the most active star-forming region in the Galaxy, the other active star-forming region that physically connects with it---i.e., Sgr\,B1---has not been explained together with Sgr\,B2 in either scenarios.
Furthermore, the CCC scenario has been gradually taken into account in recent dynamical models of the CMZ.
\citet{sor20} suggested that CCCs were naturally present in their hydrodynamic simulations, thus it is important to examine whether the observational footprints of the CCCs are consistent with those simulations and dynamical models.
Therefore, reexamination of the CCC scenario for Sgr\,B2 based on existing dynamical models is required, and this scenario also needs to be applicable mot only to Sgr\,B2 but also to Sgr\,B1.

\subsection{CCC and High-mass Star Formation}
It is well-known that a starburst in a merger is triggered by a collision between two giant molecular clouds belonging to the two galaxies \citep[see,][]{you86}.
A CCC is a mechanism that collects gas and magnetic fields at super sonic velocity.
Numerical simulations indicate that a CCC produces a compressed layer of gas that generates massive, gravitationally unstable, molecular clumps within it.
Owing to the enhancement of the magnetic field strength by shock compression, the clumps have effective Jeans masses large enough to develop into high-mass stars \citep{hab92,ino13,ino18}. 
However, it is still unknown how many new stars are triggered by a CCC.
In other words, the controlling parameters, which determine the scale of the star-forming activity initiated by a CCC, are still unknown, although they must be related to the initial physical conditions of the collision.

Thanks to recent Galactic plane surveys with 1 pc to subpc resolution---such as those done by the Mopra and Nobeyama 45 m telescopes in CO \citep{bra18,ume17}---the mechanism operating in CCCs have gradually been unveiled.
These datasets exhibit many candidate CCCs in the Galaxy; in total, more than 50 $\htwo$ regions and star clusters are possible sites of CCCs \citep[e.g.,][]{nis18,fuk18a,fuk18b,oha18,eno18,koh18,tor18,san18,hay18}.
Also, nearby extragalactic giant $\htwo$ regions, such as NGC\,604 in M\,33 and R136 in the Large Magellanic Cloud, have been suggested to be the results of large-scale collisions between two giant $\hone$ clouds \citep{tac18,fuk17}.
These results suggest that a CCC may be the major mechanism responsible for triggering high-mass star formation \citep{fuk21}.

In order to link our understanding of CCCs in starburst galaxies with those in the Galactic $\htwo$ regions, the nearest center of the galaxy---the GC of the Milky Way---is the most important target.
The GC has the highest volume density of molecular clouds in the Galaxy, and hence, it is likely to have the highest frequency of CCCs as well \citep{eno21c}.
The physical parameters of molecular clouds in the GC, such as their densities and relative velocities, are different from those in the Galactic disk.
The GC is thus a suitable region for investigating these control parameters.
As a first step for investigating CCCs in the GC, \citet{eno21a} applied a CCC identification methodology developed by \citet{fuk21} to the common footpoint of magnetic loops 1 and 2, and they found two pairs of colliding clouds.
Thus, they not only demonstrated the validity of this methodology in the GC, but also determined a unique property only seen in CCCs in the GC: they detected SiO emissions caused by a larger relative velocity of $\sim$a few tens of $\kms$.

\subsection{The Aims of This Paper}
The CCC scenario in Sgr\,B2, originally proposed by \citet{has94} and followed by \citet{sat00}, has attracted much attention.
However, there are still some uncertainties, as below.
This scenario:
\begin{enumerate}
	\item selected three velocity clouds by eye, but this is arbitrary;
	\item was only based on the distribution of molecular gas, and did not consider other wavelengths;
	\item used undersampled data sets;
	\item neglected Sgr\,B1; and
	\item did not consider the dynamical structure of the CMZ, leaving the possibility of clouds colliding at the position of Sgr\,B2 uncertain.
\end{enumerate}

In the present paper, we utilize the CCC identification method recently developed by \citet{fuk21} and \citet{eno21a}, employing recent, fully sampled data sets with large enough coverage to include Sgr\,B1 and large enough density dynamic range to investigate whether the CCC model is still applicable.
By comparing the gas distribution and recent high-resolution IR and radio-continuum data, we have constructed a multiwavelength view of the region.
We finally discuss the possibility of the CCC event and resulting star formation, together with the dynamical structures of the CMZ.

Sections 2 and 3 provide information about the methodology and data sets used in this paper.
Our main results are described in Section 4.
Section 5 is divided into three sub-Sections, in which we discuss comparisons with numerical simulations (5.1), a possible scenario based on a CCC (5.2), and comparisons with other CCCs in the Galaxy (5.3).
We summaries the present study in Section 6.

\section{Methodology}
\subsection{Signatures of a CCC}
In this Section, we first introduce the methodology developed by \citet{fuk18c,fuk21} and \citet{eno21a,eno21b} to identify a CCC.
This methodology has so far successfully identified more than 50 CCC candidates.
Based on synthetic observations of numerical simulations \citep{tak14,haw15} and on observational data of the Galactic $\htwo$ regions, \citet{fuk21} proposed the following signatures of a CCC;
\begin{enumerate}
	\item Complementary distributions between two colliding clouds: a larger cloud is hollowed out by a smaller cloud during a collision and forms the complementary spatial signature. This signature is sometimes accompanied by a spatial displacement, depending on the inclination angle of the relative motion to the LOS.
	\item A V-shaped structure or bridge feature in the position-velocity diagram: In the early stage of a CCC, the contact surface between the two clouds forms bridge features by exchanging momenta. This signature develops into a V-shaped structure when the larger cloud is hollowed out.
	\item Positional coincidence between the compressed layer and the high-mass star-forming region.
\end{enumerate}
However, not all of these signatures are always observable, because the signatures are weakened by dispersion of the natal clouds caused by the CCC and by ionization and projection effects \citep{fuk21}.
In addition, \citet{eno21a} found that CCCs in the GC are accompanied by SiO emissions, due to the larger relative velocity achieved only in the GC.
We utilize these observational signatures to identify a CCC.

\subsection{The Cold Gas Tracer}
A gas density tracer is quite important for utilizing this methodology.
Sgr\,B2 has an extremely high column density, 4 $\times$ 10$^{25}$ cm$^{-2}$ \citep{mol11}, and emissions from this region often show self-absorption.
Thus, the region is usually traced by rare molecules such as HNCO \citep[e.g.,][]{hen16}. 
On the other hand, diffuse low-density gas, which is the dominant mass reservoir, and mainly traced by CO, is also important for this work, because a CCC is a mechanism for compressing the gas, and both cause and effect---in other words, diffuse and dense gases---are required to identify a CCC.
\citet{fuk21} suggested that $\ceighteeno$ is the best tracer for the condition of Sgr\,B2.

We carefully checked various lines obtained with previous surveys, not only toward the Sgr\,B region, but also the entire CMZ, and found almost all lines, such as $\twelveco$, $\thirteenco$, SiO($J$=1--0), H$^{13}$CO$^{+}$($J$=1--0), CS($J$=1--0), HCN($J$=4--3), H$^{13}$CN($J$=1--0), HNC($J$=1--0), etc., were optically thick toward Sgr\,B2 and Sgr\,A.
On the other hand, a few lines, i. e., $\ceighteeno$, SiO($J$=5--4), HC$_{3}$N($J$=10--9), and HNCO(4--3) did not show significant self-absorption even toward Sgr\,B2.
Within these optically thin lines, SiO and HC$_{3}$N($J$=10--9) are a shock tracer and warm gas tracer, respectively, and do not show any diffuse emissions traced by CO.
Thus, they do not trace density of cold molecular gas.
By comparing a high sensitivity H$_{2}$ column density map derived from $Herschel$ data \citep{mol11}, we found $\ceighteenoh$ to be the best gas tracer, with the largest density dynamic range of density.
HNCO is a good gas tracer, also including the diffuse component but the distribution of Sgr\,B2 is apparently different from that of the H$_{2}$ column density.
This is possibly caused by the special chemical environment in Sgr\,B2(N), where rare molecules are very abundant \citep[e.g.,][]{bel13}. 
Also, the sensitivity of archival $\ceighteeno$ data was much better than that of the HNCO data.
In this paper, therefore, we mainly use $\ceighteenoh$ as the molecular-gas tracer for identifying a CCC in Sgr\,B2.

\subsection{Cloud Identification}
One of the signatures of a CCC, i. e., a V-shaped structure or bridge feature in the position-velocity diagram, as introduced above, can also be seen in the velocity gradient of a cloud.
Thus, identifying two independent clouds is important factor for the CCC methodology; whereas two clouds are often merged by the collision and are observed as a single cloud, at least partially.
We utilized a cloud separation method, using moment maps introduced by \citet{eno21b}.
This method can clearly separate two merging clouds in terms of velocity, and it has been successfully applied to GC clouds \citep{eno21a}.
The detailed justification of this method is described in the Appendix in \citet{eno21b}. 

\section{Data Sets} \label{sec:data}
\subsection{Molecular Lines}
As the main tracer of molecular gas, we use the public $\ceighteenoh$ data obtained by \citet{gin16}, taken in the on-the-fly mode with the Atacama Pathfinder Experiment (APEX). 
The spectra in the Sgr\,B region show moderate, uniform subsidences, and we subtracted them by the linear fitting of the data.
The half-power-beam-width (HPBW), spatial grid, and velocity resolution are 30\arcsec, 7\farcs2, and 1 $\kms$, respectively.

As a shock tracer, we use $\sioh$ data that was also obtained by \citet{gin16}.
We subtracted moderate, uniform subsidences from these spectra, by linear fitting of the data.
The HPBW, spatial grid, and velocity resolution for the data are also 30\arcsec, 7\farcs2, and 1 $\kms$, respectively.

In order to calculate the intensity ratio between the $J$ = 2--1 and 1--0 transitions of $\ceighteeno$, we used the $\ceighteenol$ data obtained by \citet{tok19} with the Nobeyama 45 m telescope.
These spectra also show moderate, uniform subsidences, and we subtracted them by linear fitting of the data.
The HPBW, spatial grid, and velocity resolution for the data are 21\arcsec, 7\farcs5, and 2 $\kms$, respectively.
 
\subsection{Other Wavelengths}
We mainly utilize the 90 cm radio-continuum data obtained by \citet{lar00} as an indicator of star formation activity.
In order to trace cold dust, we utilize the FIR (250, 350, and 500 $\mu$m) data obtained by the Spectral and Photometric Imaging Receiver installed on Herschel\footnote{{Herschel is an ESA space observatory with science instruments provided by European-led Principal Investigator consortia and with important participation from NASA.} \citep{pil10,gri10}.}
We also use mid-infrared (MIR; 24, 8, and 3.6 $\mu$m) data obtained with the Multiband Imaging Photometer (MIPS) and  Infrared Array Camera installed on Spitzer \citep{rie04,faz04}

\section{Results} \label{sec:res}
\subsection{Spatial Distribution of $\ceighteenoh$ toward the Sgr\,B region}

\begin{figure*}[t]
\begin{center}
\includegraphics[width=\linewidth]{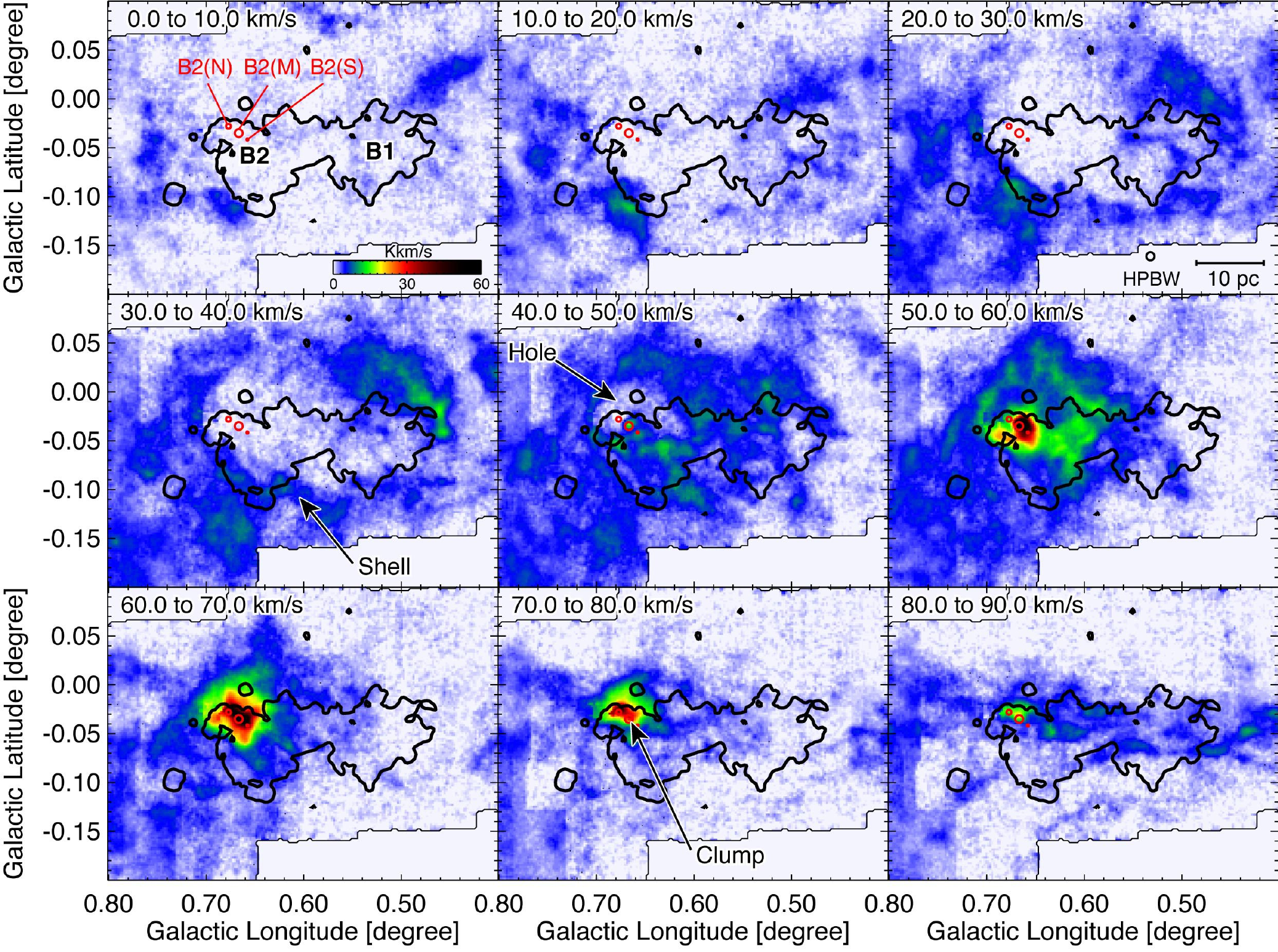}
\caption{Velocity channel distributions of $\ceighteenoh$ toward the Sgr\,B region, including Sgr\,B1 and Sgr\,B2, obtained with APEX, overlaid with thick black contours outlining the 90 cm radio-continuum emission. The red circles indicate the positions of Sgr\,B2(N), (M), and (S), respectively. Known CO features (the Shell, the Hole, and the Clump) are labeled.}
\label{lbch}
\end{center}
\end{figure*}

Figure~\ref{lbch} shows the velocity channel distribution of $\ceighteenoh$ toward the Sgr\,B region.
Molecular clouds with velocities out of the figure are located in the foreground or in the so-called ``expanding molecular ring'' \citep{kai72}, and are not associated with Sgr\,B1 and Sgr\,B2 \citep[see also Figure~4 of ][]{vas02}.
The positions of the active star-forming regions (Sgr\,B1, Sgr\,B2) and star-forming cores (Sgr\,B2(N), Sgr\,B2(M), and Sgr\,B2(S)) are indicated in the top-left panel.
The black contours indicate the boundaries of the active star-forming regions obtained from the 90 cm radio-continuum data \citep{lar00}.
As seen in Figure~\ref{lbch}, $\ceighteenoh$ traces the diffuse gas well, and it peaks at $\vlsr$ $\sim60$ $\kms$ toward Sgr\,B2(M), meaning that there is no significant self-absorption, even toward Sgr\,B2(M).
At $\vlsr$ = 20--40 $\kms$, 40--50 $\kms$, and 60--80 $\kms$, the Shell appears as a $\sim$30 pc cavity, the Hole is slightly north of Sgr\,B2, and the Clump is located near the star-forming cores, as proposed by \citet{has94}.
The figure shows that the Clump is obviously associated with the star-forming cores.
This has already been reported in the literature \citep[e.g.,][]{tsu99}.
The shapes of the Shell and the black contours agree well with each other.
The filamentary clouds elongated in the direction of Galactic longitude in $\vlsr$ $\sim$80--90 $\kms$ are probably located in another arm \citep[arm II;][]{sof95} or stream \citep{kru15}.

\subsection{Application of the CCC Identification Methodology}
In this section, we use the identification methodology developed by \citet{fuk18c} and \citet{eno21a} to investigate the possibility of a CCC in the Sgr\,B region.
As described in Section 2.2, we use the $\ceighteenoh$ line as the best tracer for the methodology in this region.

\begin{figure*}[t]
\begin{center}
\includegraphics[width=\linewidth]{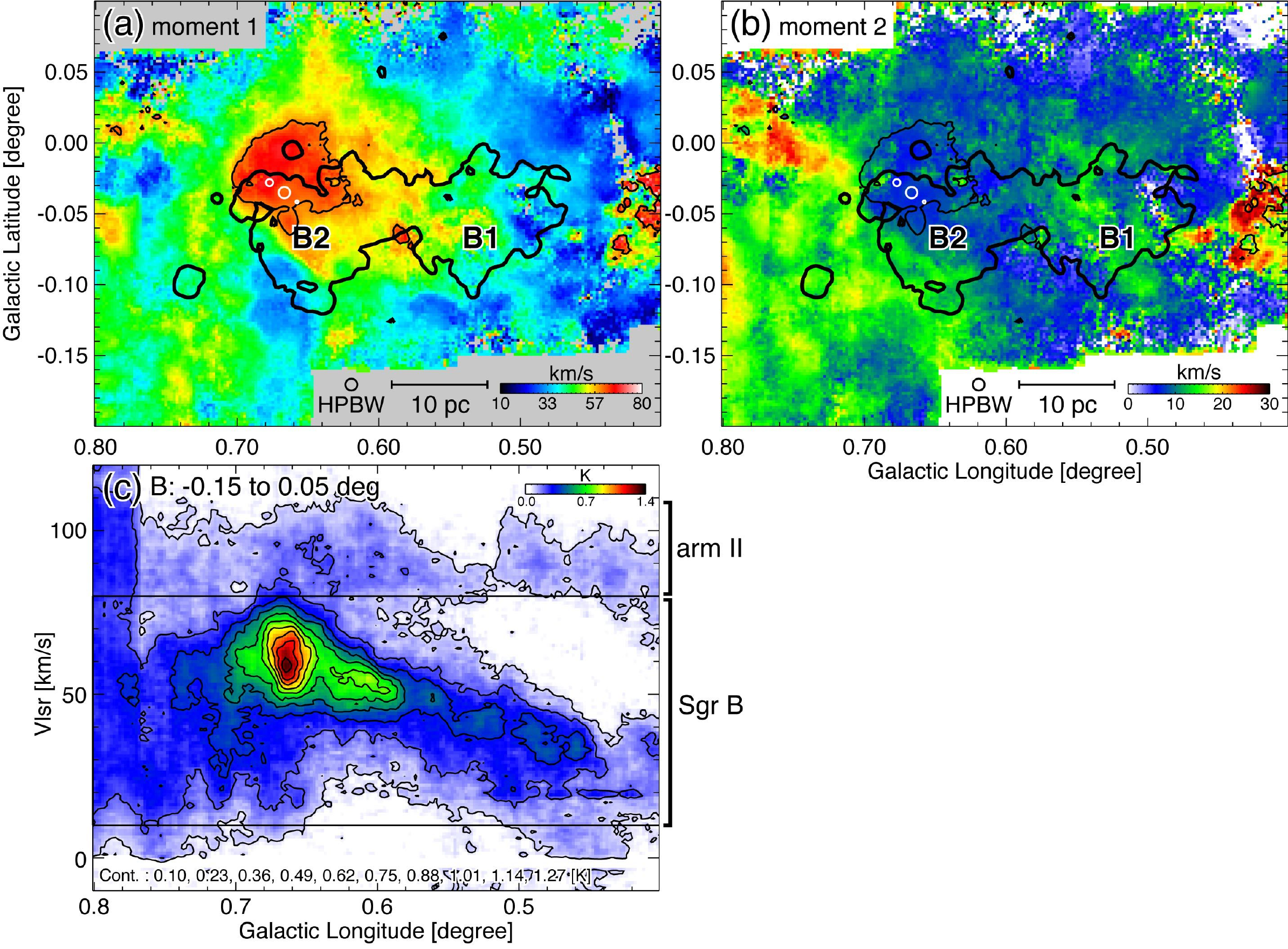}
\caption{Panels (a) and (b): intensity-weighted first-moment and second-moment maps of $\ceighteenoh$ in 10 $\le$ $\vlsr$ $\le$ 80 $\kms$. The white circles indicate the positions of Sgr\,B2(N), (M), and (S). The thin and thick black contours indicate the boundaries of the first-moment = 60.0 $\kms$ and the radio-continuum emission at 90 cm, respectively. Panel (c): longitude--velocity diagram for $\ceighteenoh$ toward the Sgr\,B region. The black lines indicate the velocity range used for the moment calculations.}
\label{moment}
\end{center}
\end{figure*}

Following \citet{eno21a}, we first define the velocity components in this region by analyzing the moments of $\ceighteenoh$.
Figures~\ref{moment}(a) and \ref{moment}(b) show the distributions of the intensity-weighted mean velocity and the standard deviation of the velocity, respectively.
We used only voxels with at least a 5 $\sigma$ significance level for this analysis, in order to reduce noise fluctuations.
Figure~\ref{moment}(c) shows a longitude-velocity diagram toward the Sgr\,B region.
The associated clouds reported in previous works are seen as major bright features in 10 $\le$ \vlsr $\le$ 80 $\kms$, indicated by the black lines.
We use this velocity range to calculate the moments.
Figure~\ref{moment}(a) shows that redshifted features (from the interior to the thin black contour: $\vlsr$ $\ge$ 60 $\kms$) are seen toward the star-forming cores in Sgr\,B2.
This component contains the Clump \citep{has94}.
We hereafter call this velocity component the ``60 $\kms$ component''.
The exterior of the black contour shows velocity gradients ranging from yellow to blue in Figure~\ref{moment}a.
We hereafter call this velocity feature the ``40 $\kms$ component''.
The eastern contact line of the 60 $\kms$ component has relatively larger line widths, colored light green (Figure~\ref{moment}(b)).

In accordance with \citet{eno21a}, we determine $\vcen$ and $\dv$, respectively, as the averages of velocity of the first and second moments in the given area, and we take the representative velocity range to be $\vcen$ $\pm$ $\dv$.
We use the interior and exterior of the thin black contour (first moment = 60.0 $\kms$) in Figure~\ref{moment}(a) as the 60 $\kms$ and 40 $\kms$ component areas, respectively.
We have confirmed that the difference between these areas does not significantly change the results shown below.
We list the representative velocity range, $\vcen$, and $\dv$ for each velocity component in Table~\ref{tab1}.

\begin{deluxetable*}{lccccc}
{\tiny
\tablenum{1}
\tabletypesize{\scriptsize}	
\tablecaption{Physical Parameters of the Velocity Components\label{tab1}}
\tablewidth{0pt}
\tablehead{
\colhead{Name} & \colhead{Representative velocity range} & \colhead{\vcen} & \colhead{$dV$} & \colhead{$\nhtwo$ (typical/peak)} & \colhead{Mass} \\
\colhead{} & \colhead{($\kms$)} & \colhead{($\kms$)} & \colhead{($\kms$)} & \colhead{($\times$10$^{23}$ cm$^{-2}$)} & \colhead{($\times$10$^{6}$ \msun)}
}
\decimalcolnumbers
\startdata
40 $\kms$ component& 29 -- 53 & 41.6& 12.2  & $\sim$0.5 / 1.7 & 1.4\\
60 $\kms$ component& 53 -- 74 & 63.8& 10.4  & $\sim$1 / 6.6 & 1.1\\
\enddata
\tablecomments{Column 1: names of the components, column 2: velocity ranges; column 3: peak velocities derived from the moment 1 map; column 4: velocity line widths derived from the moment 2 map; column 5: typical and peak molecular column densities toward each component; column 6: molecular masses.}
}
\end{deluxetable*}

\begin{figure*}[t]
\begin{center}
    \includegraphics[width=12cm]{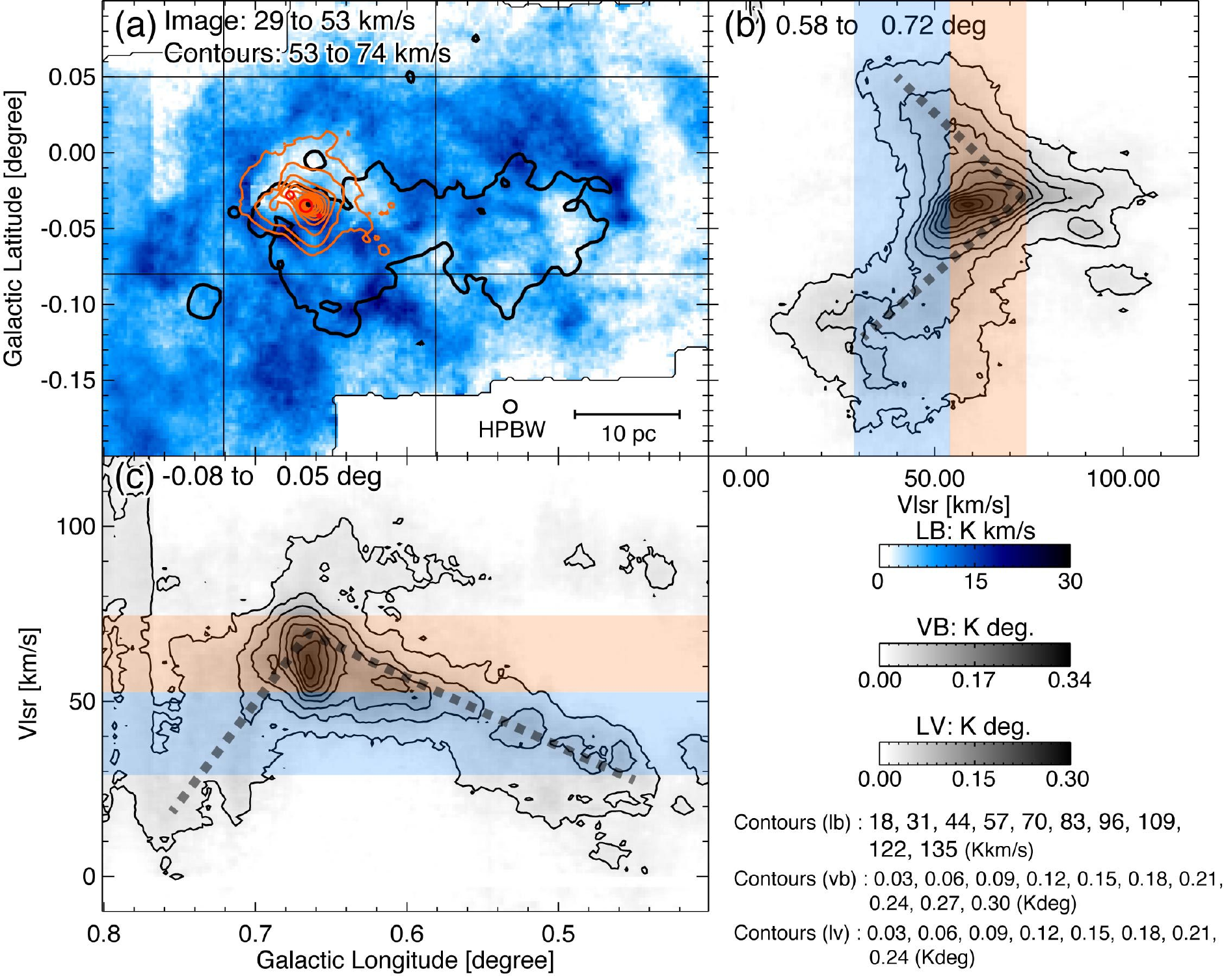}
  \caption{Panel (a): integrated intensity distributions of the 40 $\kms$ component (blue color) and 60 $\kms$ component (orange contours) in $\ceighteenoh$. Panels (b)and (c): velocity--latitude diagram and longitude--velocity diagram of $\ceighteenoh$ toward the Sgr\,B region. The velocity integration ranges for the 40 $\kms$ and 60 $\kms$ components are indicated by the transparent blue and orange belts, respectively.}
  \label{lbv}
  \end{center}
\end{figure*}

Figures~\ref{lbv}(a)--\ref{lbv}(c) show the integrated intensity distributions, the velocity--latitude diagram, and the longitude--velocity diagram for the 40 $\kms$ and 60 $\kms$ components in $\ceighteenoh$.
The two velocity components exhibit the typical signature of a CCC: a complementary distribution in space (Figure~\ref{lbv}(a)).
The star-forming cores, indicated as the red circles, are located at the intensity peak of the higher--column density cloud, the 60 $\kms$ component.
Furthermore, there are rotated, V-shaped features in the position--velocity diagrams.
All of these features are typical signatures of a CCC.
We also investigated the SiO emissions and the line-intensity ratios between $\ceighteenoh$ and $\ceighteenol$ (\rco) toward this region (see the Appendix, Figures~\ref{sio} and \ref{ratio}).
From Figure 4 of \citet{nis15}, the typical value of $\rco$ and those in the star-forming regions in Orion are $\sim$0.5 and 1.0--1.2, respectively.
In the Sgr\,B region, a typical value of $\rco$ is $\sim$0.8; it is higher at the boundary of the Shell and the boundary of the Hole (1.0--1.6) and highest at the Clump (1.6--2.8).
The $\sioh$ emissions are detected only in the areas where $\rco$ is higher.

\subsection{Physical Parameters of the Two Components}
Here, we derive physical parameters, such as column densities and masses, for the 40 $\kms$ and 60 $\kms$ components, assuming local thermodynamic equilibrium (LTE).
We assume that the excitation temperature is $\tex$ = 30 K, which is similar to the temperatures of previous studies \citep{has94,mor83}.
Although the temperatures may vary with position or component, the differences in $\tex$ do not significantly affect the results.

First, we define the equivalent brightness temperature $J(T)$ as
\begin{equation}
  J(T) = \frac{h\nu}{\kb}~[~\exp(\frac{h\nu}{\kb T - 1})~]^{-1}.
\end{equation}
where $h$, k$_{B}$, and $\nu$ are Planck's constant, Boltzmann's constant, and the frequency, respectively.
A variation of the radiative transfer equation gives the optical depth $\tau_{\nu}$ as
\begin{equation}
  \tau (v) = -\ln(~1~-~\frac{\tobs}{J(\tex)~-~J(\tbg)}~).
\end{equation}
where $\tobs$ and $\tbg$ are the observed intensity on the $\tmb$ scale and the temperature of the background (= 3 K), respectively.
The column density $N$ of the molecules can be calculated as follows:
\begin{equation}
  N = \sum_{V} ~\tau (v)~\Delta v~\frac{3\kb\tex}{4\pi^{3}\nu\mu^{2}}~\exp(\frac{h\nu J}{2\kb\tex}) ~\frac{1}{1 - \exp({-\frac{h\nu}{\kb\tex}})},
\end{equation}
where $V$, $\Delta v$, $\mu$, and $J$ are the velocity of the cloud, the velocity resolution of the data, the electric-dipole moment of the molecule, and the rotational transition level, respectively. 
Substituting $\kb$ = 1.38 $\times~10^{-16}$ (erg/K), $\tex$ = 30 (K), $\nu$ = 2.196 $\times~10^{11}$ (Hz), $\mu$ = 1.10 $\times~10^{-19}$ (esu~cm), $h$ = 6.63 $\times~10^{-27}$ (erg~s), and $J$ = 1 into Equation (3) gives
\begin{equation}
  \nco = 1.518 \times 10^{16} ~\sum_{V} ~\tau (v)~\Delta v.
\end{equation}
The relation between $\nco$ and $\nhtwo$ was determined by \citet{fre82} to be
\begin{equation}
  \nhtwo = (\frac{\nco}{1.7 \times 10^{14}} + 1.3) \times 10^{21}.
\end{equation}
 
Using Equations (4) and (5), we find the typical / peak column densities of the 40 $\kms$ and 60 $\kms$ components to be 0.5 / 1.7 $\times$ 10$^{24}$ and 1 / 6.6 $\times$ 10$^{23}$ cm$^{-2}$, respectively.

The molecular mass is determined from following equation:
\begin{equation}
  M = \mum \mpro \sum_{i} ~[d^2 \Omega \nhtwo_{,i}],
\end{equation}
where $\mum$, $\mpro$, $d$, $\Omega$ and $\nhtwo_{,i}$ are the mean molecular weight, proton mass, distance, solid angle subtended by a pixel, and column density of molecular hydrogen for the $i$th pixel, respectively.
We assume a helium abundance of 20$\%$, which corresponds to $\mum$ = 2.8.
If we take $d$ = 8300 pc, the masses of the 40 $\kms$ and 60 $\kms$ components are $\sim$1.4 and $\sim$1.1 $\times$ 10$^6$ $\msun$, respectively.
These physical parameters are listed in Table~\ref{tab1}.

\subsection{Multiwavelength View of the Sgr B Region}
Above, we have defined two velocity components (the 40 $\kms$ component and 60 $\kms$ component), and we have shown that a CCC model is applicable in this region.
In this section, we present a multiwavelength view of the region and compare it with the results obtained in the previous section.

\begin{figure*}[t]
\begin{center}
    \includegraphics[width=15cm]{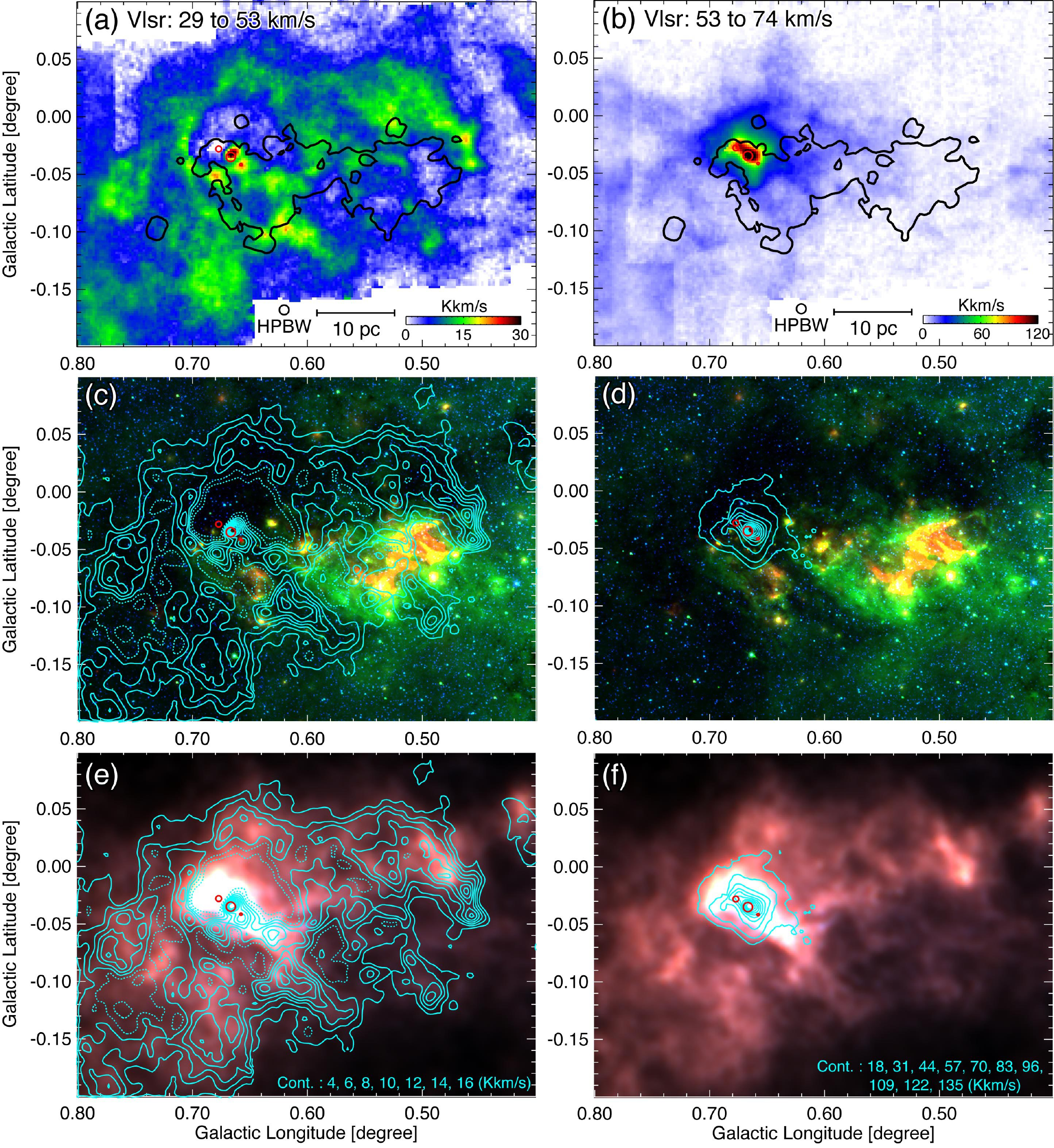}
  \caption{Panels (a) and (b): integrated intensity distributions of the 40 $\kms$ component and the 60 $\kms$ component in $\ceighteenoh$, obtained with APEX. The red circles indicate the positions of Sgr\,B2(N), (M), and (S), from north to south. The thick black contours indicate the outline of the radio-continuum emission at 90 cm. Panels (c) and (d): the cyan contours obtained from the distributions of  $\ceighteenoh$ in Figures \ref{multi}(a), and \ref{multi}(b), respectively are superposed on a three-color composite image obtained with Spitzer. The red, green, and blue correspond to 24, 8, and 3.6 $\mu$m, respectively. Panels (e) and (f): the same contours of Figures~\ref{multi}(b), and \ref{multi}(c), respectively, are superposed on a three-color composite image obtained with Herschel. Here, the red, green, and blue correspond 500, 350, and 250 $\mu$m, respectively.}\label{multi}
  \end{center}
\end{figure*}

Figure~\ref{multi}(a) shows the spatial distribution of the 40 $\kms$ component, overlaid with the black contours of the boundary of the star-forming region traced by the 90 cm radio-continuum emission.
The representative velocity ranges defined in Section 4.2 are used as the integration ranges.
Figures~\ref{multi}(c) and \ref{multi}(e) are superpositions of contours of the contours of the 40 $\kms$ component on the MIR and FIR images obtained with Spitzer and Herschel, respectively.
The FIR composite image shows only a single burgundy color.
This likely indicates that the properties of the cold dust emitting the FIR are the same throughout this region, except for the white area, where the radiation is very strong and saturated.
The cold dust is seen as dark clouds in the MIR, which means that a bright MIR source radiates from behind.
The major radiative process in the MIR is irradiation with UV photons from high-mass stars, so the MIR is another indicator of star-forming activity, in addition to the radio-continuum.
It is clear that the bright MIR regions coincide well, excluding the vicinity of the star-forming cores, with the radio-continuum boundary (the black contour in Figure~\ref{multi}(a)).
Thus, the bright background sources are Sgr\,B1 and Sgr\,B2.

The distribution of the 40 $\kms$ component in the dense-gas tracer, $\ceighteenoh$, is very similar to that of the cold dust or the MIR dark clouds.
These three therefore originate from the same object at different wavelengths, and they are located in front of Sgr\,B1 and Sgr\,B2.
At the northwest edge of Sgr\,B1, the integrated intensity of $\ceighteenoh$ is enhanced and well delineate the boundary of the radio-continuum emission (Figure~\ref{multi}(a)); thus, the 40 $\kms$ component is likely interacting with Sgr\,B1.
These are the dense clouds labeled ``cloud d'' and ``cloud e'' by \citet{lon13}.

Figure~\ref{multi}(b) also shows the spatial distribution of the 60 $\kms$ component in $\ceighteenoh$, overlaid with the black contours of the boundary of the star-forming region traced by the 90 cm radio-continuum emission. 
Figures~\ref{multi}(d) and \ref{multi}(f) are similar superpositions of the contours of the 60 $\kms$ component on the MIR and FIR images obtained with Spitzer and Herschel, respectively.
The 60 $\kms$ component coincides with the strong FIR peak and the star-forming cores in Sgr\,B2.
This result clearly shows that the star-forming cores of Sgr\,B2 are still at a very young stage of cluster formation, and are deeply embedded in the high-density cloud with $\nhtwo$ $\ge$ 10$^{23}$ cm$^{-2}$, as already reported in ealier studies.
The 60 $\kms$ component is also dark in the MIR image, although this region is very bright in the radio-continuum. Thus, the 60 $\kms$ component is located a bit closer to the Sun than Sgr\,B2.
The darkness in the MIR image also indicates that Sgr\,B2 is a much younger system than Sgr\,B1.

\subsection{Summary of Observational Insights}
Here, we summarize the observational aspects of the Sgr\,B region and the results of our analyses.
\begin{enumerate}
\renewcommand{\labelenumi}{\roman{enumi})}
	\item Following \citet{eno21a}, we identified two velocity features (the 40 $\kms$ and 60 $\kms$ components) associated with Sgr\,B1 and Sgr\,B2 by using moment maps of $\ceighteenoh$.
	\item These two components clearly show all the signatures of a CCC proposed by \citet{fuk18c}: the complementary spatial distribution between the two components, the V-shaped structure formed by the two colliding clouds in the position--velocity diagrams, and the coincidence between the location of the intensity peak of the denser component and the high-mass star-forming region. We also detected SiO emission, which accompanies CCCs in the GC \citep{eno21a}.
	\item Using the LTE approximation, we estimated the physical parameters of the two velocity components with $\ceighteenoh$. The peak column densities and masses of the 40 $\kms$/60 $\kms$ components are $\sim$2 $\times$ 10$^{23}$ cm$^{-2}$ / $\sim$7 $\times$ 10$^{23}$ cm$^{-2}$, and 1.4 $\times$ 10$^6$ $\msun$ / 1.1 $\times$ 10$^6$ $\msun$, respectively.
	\item By comparing the CO distributions and other wavelengths, we found that the 40 $\kms$ component, which is observed as an IR dark cloud, is located in front of Sgr\,B1 and Sgr\,B2. The 60 $\kms$ component is a high-density dust clump, with $\nhtwo$ $\ge$ 10$^{23}$ cm$^{-2}$, which includes star-forming cores. It is at a very young stage of cluster formation, and is located in front of Sgr\,B2.
\end{enumerate}

\section{Discussion} \label{sec:disc}
In the previous Section, we applied the CCC identification methodology to the Sgr\,B region and found clear signatures of a CCC.
In this section, we investigate further observational evidence for a CCC and discuss the CCC scenario, including star-forming activities.

\subsection{Comparison with Numerical Simulations}
\begin{figure*}[t]
\begin{center}
    \includegraphics[width=\linewidth]{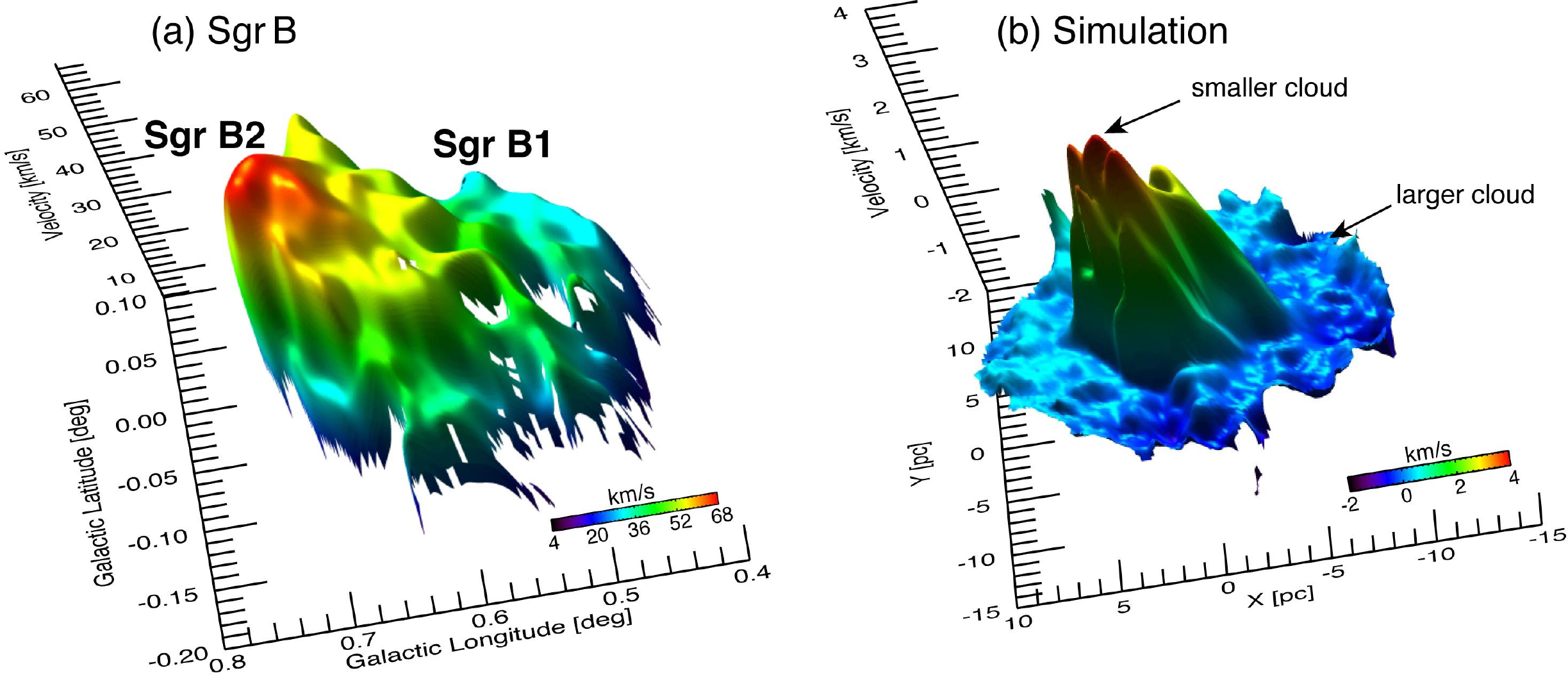}
  \caption{Panel (a): longitude-latitude-velocity surface plot of the Sgr\,B region in $\ceighteenoh$. Panel (b): ppv surface plot of a numerical simulation of a CCC performed by \citet{tak14} and \citet{haw15}.}
  \label{surface}
  \end{center}
\end{figure*}

Figures~\ref{surface}(a) and \ref{surface}(b), respectively, show a lngitude-latitude-velocity surface plot of $\ceighteenoh$ in the Sgr\,B region and a position-position-velocity (ppv) surface plot in the $\twelvecol$ synthetic observational data obtained by \citet{haw15} (for a head-on collision with a relative velocity of 10 $\kms$, observed from the angle of the relative motion).
The color in Figure \ref{surface}(a) corresponds to velocity.
Both ppv plots show similar, cone-shaped structures.
\citet{hen16} found a similar structure by using HNCO data obtained with the Mopra telescope \citep{jon12}.
Note that these surface density plots were produced by velocities derived from the first moment (intensity-weighted mean velocity), and hence the velocity of the smaller cloud in the plot reflects the velocities of both the smaller and larger clouds, and does not directly present its original velocity.
Thus, the quantitative comparison between Figure~\ref{surface}(a) and \ref{surface}(b) does not make sense.

The conical structures seen in Figure~\ref{surface} typically result from a collisional interaction between two clouds, due to the exchange of the clouds' momenta, which is observed as the V shape in the position--velocity diagram (See Section 2.1).
This structure is easily weakened by projection effects \citep[see][for details]{fuk18c}.
The observation shows a clear conical structure, but the apex is shifted slightly from the center, in the direction of positive Galactic longitude, suggesting that the collision may not be head-on and that the 60 $\kms$ component has a relative velocity toward positive Galactic longitudes \citep[see][for details]{fuk18c}.

The part of the larger cloud that surrounds the cone shape, but that does not participate in any collisional interactions, preserves the larger cloud's original velocity structures, as illustrated by the light and dark blue in Figure~\ref{surface}(b).
In the observational data, the velocity gradient of the cone extends from the eastern boundary of Sgr\,B2 to the western boundary of Sgr\,B1.
Thus, if we adopt the CCC model, a comprehensive CCC probably occurred between the two clouds, as shown in light blue and red in Figure~\ref{surface}(a).
These two clouds correspond to the Shell and the Clump.
Sgr\,B1 is located in the interior of the Shell, and thus it is possible that not only Sgr\,B2, but also Sgr\,B1, were triggered by the CCC.
Larger velocity widths are observed at the collisional interface (that is, at the base of the cone structure in the synthetic observations), and the enhanced line widths and SiO emissions detected along the inner boundary of the Shell lend support to a comprehensive collision at the 30 pc scale.

Figure~\ref{multi}(c) shows that the 40 $\kms$ component, which appears dark, intercepts the MIR emission from the background source, Sgr\,B1.
In spite of this, the boundary of the MIR emission from Sgr\,B1 coincides with the optically thin radio-continuum emission, which is not intercepted by the foreground cloud of the 40 $\kms$ component (see Figure~\ref{multi}(a)).
This means that the star-forming activities of Sgr\,B1 only spread toward the region where the CO intensity of the 40 $\kms$ component is depressed.
This is clear evidence that a past collision at the position of Sgr\,B1 hollowed out the 40 $\kms$ component and created the hole, and that the collision triggered high-mass star formation in Sgr\,B1.
Conversely, if we abandon the CCC model, the coincidence between the CO and radio-continuum toward Sgr\,B1 is rather unnatural.

\subsection{CCC Scenario}
\begin{figure*}[t]
\begin{center}
    \includegraphics[width=13cm]{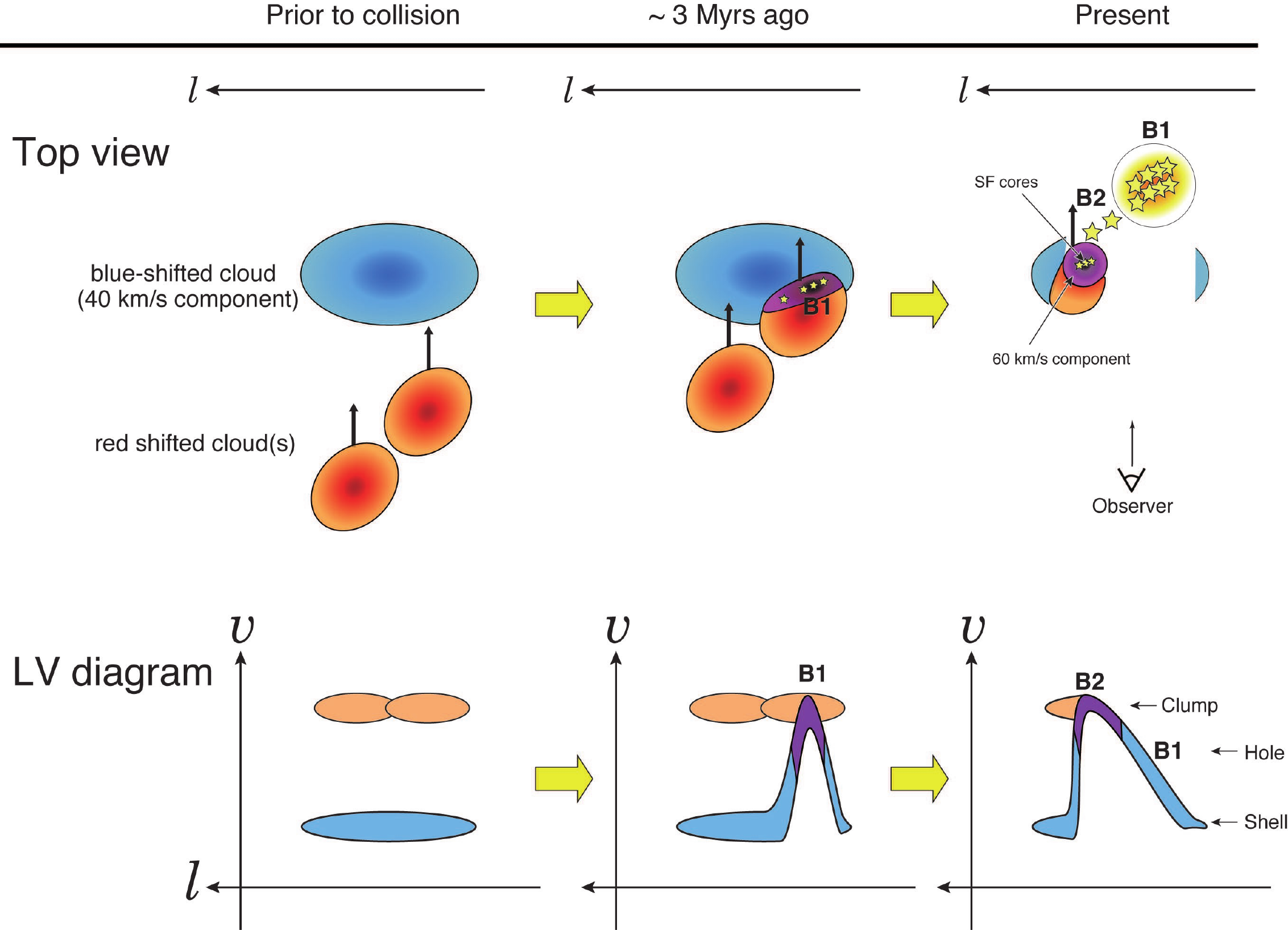}
  \caption{Upper panel: schematic image of the time evolution of the CCC in the Sgr\,B region (top view). Lower panel: schematic image of the time evolution of the longitude--velocity diagram.}
  \label{sche}
  \end{center}
\end{figure*}

From the above discussion, we here propose the following plausible collision scenario: two redshifted clouds---or one large redshifted cloud with two density peaks---approached a blueshifted cloud (the leftmost panel of Figure~\ref{sche}).
About 3 Myr ago, the two velocity features started to collide at the position of Sgr\,B1, and formed Sgr\,B1 first (the middle panel of Figure~\ref{sche}).
The region involved in this collision then extended from west to east (from Sgr\,B1 to Sgr\,B2).
A recent collision at the positions of Sgr\,B2(N), (M), and (S) triggered star-forming activities there (the rightmost panel of Figure~\ref{sche}).
The blueshifted cloud is likely to be the 40 $\kms$ component.
If we assume the age of Sgr\,B1 to be 3 Myr, then, by dividing the distance between Sgr\,B2 and Sgr\,B1 by the age, we obtain the speed at which the collision spread along the Galactic longitude as 30 pc / 3 Myr = 10 $\kms$.

We suggest two possible candidates for the redshifted cloud(s).
One possibility is that the clouds are located in arm II at $\vlsr$ $\sim$90 $\kms$, as shown in Figures~\ref{lbch} and \ref{moment}(c).
These figures show that the 60 $\kms$ component connects to arm II in space and velocity.
In this case, the velocity difference of $\sim$30 $\kms$ between the 60 $\kms$ component and arm II can be explained by the exchange of momenta between the 40 $\kms$ component and arm II (90 $\kms$).
The other possibility is a cloud with an initial velocity of $\sim$60 $\kms$.
If this cloud had a very high-density, compared to that of the 40 $\kms$ component, the cloud might maintain its initial velocity even after the collision.
However, there is so far no clear evidence for the existence of clouds with 60 $\kms$ in this direction, and furthermore, the 60 $\kms$ component has apparently extended gas to the higher velocity (see e.g., Figure~\ref{lbv}(b)), then the first case might be more plausible.

According to \citet{fuk18c} two velocity components are not always observable in a CCC, due to projection effects and so on.
\citet{fuk18b} present a case of a CCC with a single-velocity component in the star cluster GM24.
Therefore, even though the conical structure in the ppv plot consists of a single-peak velocity profile, this model is still possible.

\begin{figure*}[t]
\begin{center}
    \includegraphics[width=13cm]{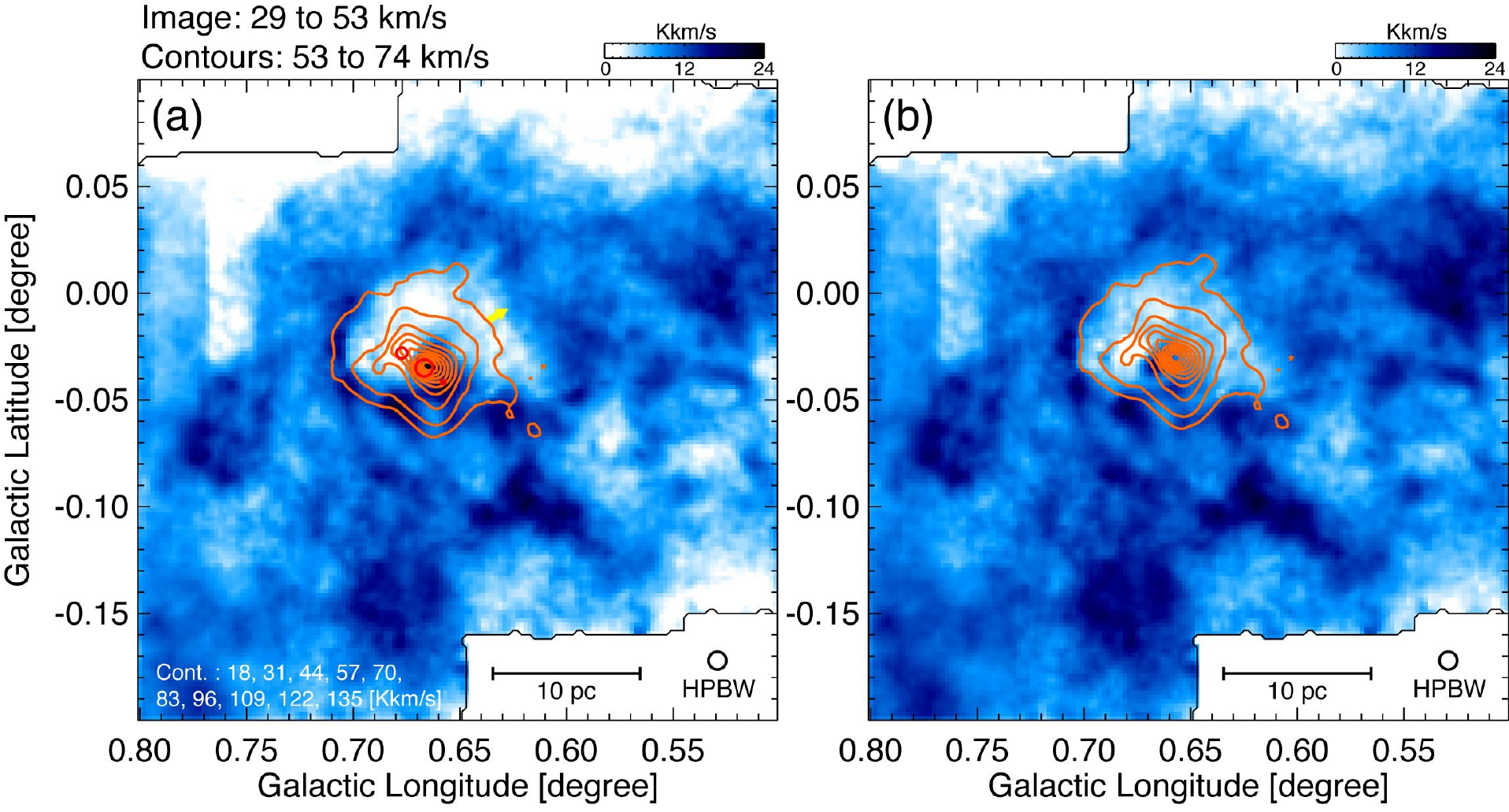}
  \caption{Panel (a): close-up view of integrated intensity distribution of the 40 $\kms$ component (blue color) and the 60 $\kms$ component (orange contours). The red circles indicate the positions of Sgr\,B2(N), (M), and (S). Panel (b): the same as Figure~\ref{displace}(a), but the contours are displaced 4 pixels (1.16 pc) to the west and 2 pixel (0.58 pc) to the north. The length and the direction of the displacement is indicated by the yellow arrow in Figure~\ref{displace}(a).}\label{displace}
  \end{center}
\end{figure*}

Figure~\ref{displace}(a) shows distributions of the 40 $\kms$ component (blue contours) and the 60 $\kms$ component (orange contours).
The distributions of the hole in the 40 $\kms$ component and the 60 $\kms$ component display a slight displacement.
Following \citet{eno21a}, and by using an algorithm developed by Fujita et al. (2021), we estimate the displacement to be $\sim$1.3 pc to the northwest, which maximizes the complementarity between the two components.
The yellow arrow in Figure~\ref{displace}(a) indicates the derived displacement vector.
Figure~\ref{displace}(b), which shows the contours displaced by that arrow overlaid on the 40 $\kms$ component, exhibits perfect coincidence between the two components.
Three active star-forming cores are located at the center of the 60 $\kms$ component.
If we assume that the LOS relative velocity and the relative velocity in the direction of Galactic longitude are same, then by dividing the displacement by the relative velocity, we estimate the duration time of the collision at the star-forming cores to be 1 pc/30 $\kms$ $\simeq$ 4 $\times$ 10$^{4}$ yr.
This value is comparable to the typical age of hypercompact $\htwo$ regions \citep[10$^3$--10$^{4}$ yr, e.g.,][]{qin08}.
From numerical MHD simulations, \citet{fuk21} reported that molecular flows colliding at a relative velocity of 20 $\kms$ and with densities of 1000 cm$^{-3}$ generate massive clumps at the collision interface, with mass accretion rates larger than 10$^{-4}$ $\msun$ yr$^{-1}$.
Since the 60 $\kms$ component is denser than their simulations, a CCC is a viable process for creating the star-forming cores.

\subsection{Origin of the Large Velocity Difference between the Two Clouds}
The driving force responsible for producing the large relative velocity of a few tens of $\kms$ between two massive clouds is still unclear, but we propose two possibilities.
One is a bar-potential-driven model.
As \citet{bin91} originally proposed, two families of x1 and x2 orbits can be produced in the central 1 kpc of a galaxy under certain bar-potential conditions.
The probability of a CCC increases at the intersections of these orbits (or arms).
According to the x1--x2 orbits dynamical model, as in \citet{bin91}, \citet{bal10}, and \citet{mol11}, the Sgr\,B region is located at one of the intersections.

Based on hydrodynamic simulations, \citet{sor19} discussed the origins of extended velocity features (EVFs), and suggested that EVFs are consequently obtained as results from two types of cloud collisions, overshot type or connection type, accelerated by the bar potential.
The overshot type requires an interloper, which is overshoot from the dust lane.
The redshifted clouds likely belong to arm II (see Section 5.2), and have not been overshoot; thus, the other cloud---i.e., the blueshifted cloud, or, in other words, the 40 $\kms$ component---should be overshot from a dust lane resulting from bar-potential acceleration, if this model is true.
Although the mass and size of the overshoot cloud is quite large (see Table1), and thus it is apparently unnatural, it still remains a possible scenario.
If the Sgr B region is a connection type EVF, it needs to be a connecting point for at least two streams.
As discussed at the beginning of this section, this scenario is also possible.

Another possibility is a magnetically driven model.
The GC has strong magnetic fields at least 50 $\mu$G \citep{cro10}, and such conditions can easily excite Alfv\'en velocities of the order of a few tens of $\kms$ \citep[see][]{suz15}.
Some authors have also reported observational evidence of loop-like gas streams perpendicular to the Galactic plane produced by magnetic buoyancy \citep{fuk06,eno14}.
The probability of a CCC increases at the footpoints of these loops.
\citet{eno21a} discovered an enhancement in the probability of a CCC at the footpoints of the loops.
They suggested that the enhancement is caused by rear end collisions originated from the stagnation of the flow at the footpoints. 
According to the open-orbit dynamical model, the orbit has warps in $z$-direction.
Although this warp has so far not been clearly explained, if we assume that the warp is reflecting the shape of toroidal magnetic field line, the Sgr\,B region is located at the foot point of the loop (i. e., the base of the warp).
Thus, the collision at the Sgr\,B region can be interpreted as a rear end collision resulting from the gravitationally accelerated cloud sliding down along the loop.

\subsection{The Star Formation Triggered by the CCC Event}
\begin{figure*}[t]
\begin{center}
    \includegraphics[width=9cm]{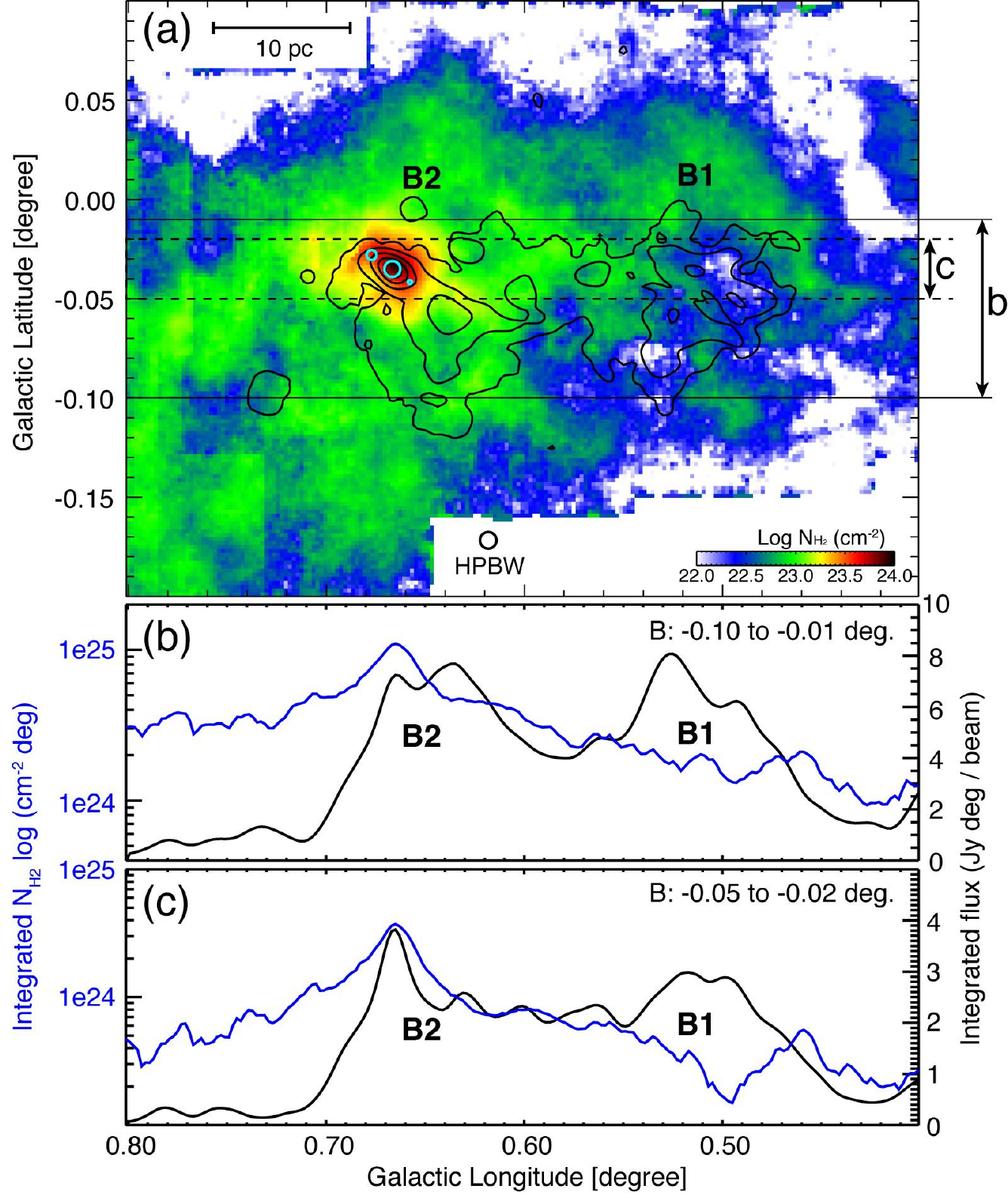}
  \caption{Panel (a): H$_{2}$ column-density map of the molecular gas associated with the Sgr\,B region estimated from the $\ceighteenoh$ emission in 20 $\kms$ $\le$ $\vlsr$ $\le$ 90 $\kms$. The contours show the distribution of the radio-continuum emission at 90 cm. The cyan circles indicate the positions of Sgr\,B2(N), (M), and (S). The integration latitude ranges for Figures~\ref{sf}(b) and \ref{sf}(c) are indicated by the black lines and dashed black lines, respectively. Panel (b): the integrated column density (blue) and integrated 90 cm flux (black) plots toward the star-forming regions in the Sgr\,B region. Panel (c): the same plots for the limited latitude range focused on the Sgr\,B2 star-forming cores.}
  \label{sf}
  \end{center}
\end{figure*}

The image and contours in Figure~\ref{sf}(a) show the distributions of the column densities ($\nhtwo$) derived by using the LTE approximation and the radio-continuum emission, respectively.
We adopted $\tex$ = 30 K and the velocity range 20 $\le$ $\vlsr$ $\le$ 90 $\kms$ for the calculation.
Figure~\ref{sf}(a) clearly shows that the column density is typically $\sim$10$^{23}$ cm$^{-2}$ and increases up to $\sim$10$^{24}$ cm$^{-2}$ toward the star-forming cores, and it decreases to less than $\sim$10$^{22}$ cm$^{-2}$ toward Sgr\,B1.
We assume here that the radio-continuum emission is optically thin, in which case it is a good tracer of star formation activity.
Figures~\ref{sf}(a) shows that the correlations between gas and star-formation activity get better/worse toward Sgr\,B2/Sgr\,B1.

The plots of the integrated $\nhtwo$ and radio-continuum flux versus the Galactic longitude shown in Figures~\ref{sf}(b) and \ref{sf}(c) delineate the above picture more quantitatively.
The integrated latitude ranges of Figure~\ref{sf}(b) and \ref{sf}(c) correspond to the two star-forming regions and the star-forming cores, respectively.
Figure~\ref{sf}(b) shows that the star-forming activities in Sgr\,B1 and Sgr\,B2 are fairly comparable, although the mass of the molecular gas concentrated in Sgr\,B2 is one order of magnitude higher than in Sgr\,B1.
This may mean that 90$\%$ of the molecular gas in Sgr\,B1 has been consumed to form stars or has been dissipated by ionization or the CCC. 
As discussed in Section 5.1, the shape of the cavity in the 40 $\kms$ component and Sgr\,B1 are very similar, although they have slightly different LOS distances.
Therefore, the collision may have hollowed out the 40 $\kms$ component, creating the cavity and triggering high-mass stars, thus dissipating the natal cloud.
Figure~\ref{sf}c shows that $\sim$50$\%$ of the star-forming activity and the molecular gas in Sgr\,B2 are contained in a small volume.
On the other hand, the molecular gas in Sgr\,B1 is significantly decreased.
Again, this depression is likely caused by the collision and ionization.

\subsection{Position of the Sgr B Region among Other CCC Regions}
\citet{eno21a} collected all the literature reporting the signatures of CCCs, and they carried out a statistical analysis of the data.
For more than 50 Galactic objects, they found correlations between the number of O- and early B-type stars and the peak column densities of the colliding clouds, as well as between the relative velocities and peak column densities of the colliding clouds \citep[see Figure~9 of ][]{eno21a}.
The CCC in the Sgr\,B region shows the following relative velocity, peak column density, and number of high-mass stars (candidates): 22 $\kms$, $\sim$10$^{24}$, and a few hundreds, respectively, in the star forming cores.
These physical parameters mostly satisfy the correlations found by \citet{eno21a} for the most massive cases.

The correlation plots indicate that only a CCC with a lower column density triggers the formation of a normal star cluster, that a moderate column density triggers a superstar cluster, and that a high column density may trigger a starburst.
The threshold column densities for triggering a normal cluster, a superstar cluster, and a starburst are $\sim$10$^{21.5}$, $\sim$10$^{23}$, and $\sim$10$^{24}$ cm$^{-2}$, respectively.
Some authors have suggested that a collision between two large-scale $\hone$ flows can trigger a massive cluster containing M$_{*}$ of a few hundred thousands $\msun$, such as R136 in the Large Magellanic Cloud \citep{fuk17,tac18}.
Detailed molecular observations of the star-forming cores in Sgr\,B2 with ALMA may reveal the CCC star-forming process at a very young stage, and link the triggering mechanism with the active star formation seen in the external galaxies.

\section{Summary} \label{sec:sum}
We have used a recently developed CCC identification methodology to investigate the possibility of a CCC in the Sgr\,B region, by using datasets at various wavelengths and fully sampled molecular-line data.
Our major findings are summarized as follows.
\begin{enumerate}
	\item We identified two CO components, with velocities of $\sim$42 (the 40 $\kms$ component) and $\sim$64 $\kms$ (the 60 $\kms$ component) by using $\ceighteenoh$ as the optically thin, dense-gas tracer. These velocity components form a cone-like structure covering both Sgr\,B1 and Sgr\,B2 in the ppv plot.
	\item From comparisons with other wavelengths, we identified the locations of the velocity components and the star-forming regions in the Sgr\,B region as follows: Sgr\,B1 and Sgr\,B2 are located farthest from the Sun, while the 40 $\kms$ component is nearest, and intercepts the light from Sgr\,B1 and Sgr\,B2. The star-forming cores in Sgr\,B2---namely Sgr\,B2(N), (M), and (S), accompanied by the 60 $\kms$ component---are located nearer to the remaining region of Sgr\,B2. According to previous work, Sgr\,B2 is likely to be located nearer to the Sun than Sgr\,B1.
Although the LOS distances to Sgr\,B1 and the 40 $\kms$ component are likely to be slightly different, the distributions of these two regions show an interaction between them.
	\item Two clouds satisfy all the signatures of a CCC, as suggested by \citet{fuk18c}. If we apply the CCC model, Sgr\,B1 and Sgr\,B2 are explained as results of contiguous collisions between the 40 $\kms$ and redshifted clouds. One of the cores of the redshifted cloud, with a column density of 10$^{23-24}$ cm$^{-2}$, collided first and formed Sgr\,B1 a few Myr ago, while a very recent collision of another core of the redshifted cloud a few $\times$ 10$^4$ yr ago triggered active star formation in the star-forming cores in Sgr\,B2, which include a few hundred high-mass stars. The candidates for the redshifted cloud are either a cloud with a velocity of $\sim$60 $\kms$ or the so-called arm II.
	\item The CCC in the Sgr\,B region satisfies the correlation among the relative velocity, peak column density, and the number of O- and early B-type stars found in \citet{eno21a} in the most active star formation case. 
\end{enumerate}

Although we do not have clear evidence for this, the mechanism responsible for driving the large relative velocity of the clouds may either be an encounter between arms or streams, or else a magnetic instability, such as the Parker instability.
Higher-resolution observations will be required in order to determine the detailed mechanism responsible for triggering star formation in Sgr\,B2.

\acknowledgments
We thank the anonymous referee(s) for helpful comments that improved the manuscript.
RE is grateful to Dr. T. Hasegawa for helpful discussions and comments on this work. The authors would like to thank Enago (www.enago.jp) for the English language review.
This work is based [in part] on observations made with the Spitzer Space Telescope, which is operated by the Jet Propulsion Laboratory, California Institute of Technology under a contract with NASA.
This work was financially supported Grants-in-Aid for Scientific Research (KAKENHI) of the Japanese society for the Promotion for Science (JSPS; grant number society for 20K14520).

\vspace{5mm}


\appendix
\section{Appendix information}
\begin{figure*}[t]
\begin{center}
    \includegraphics[width=16cm]{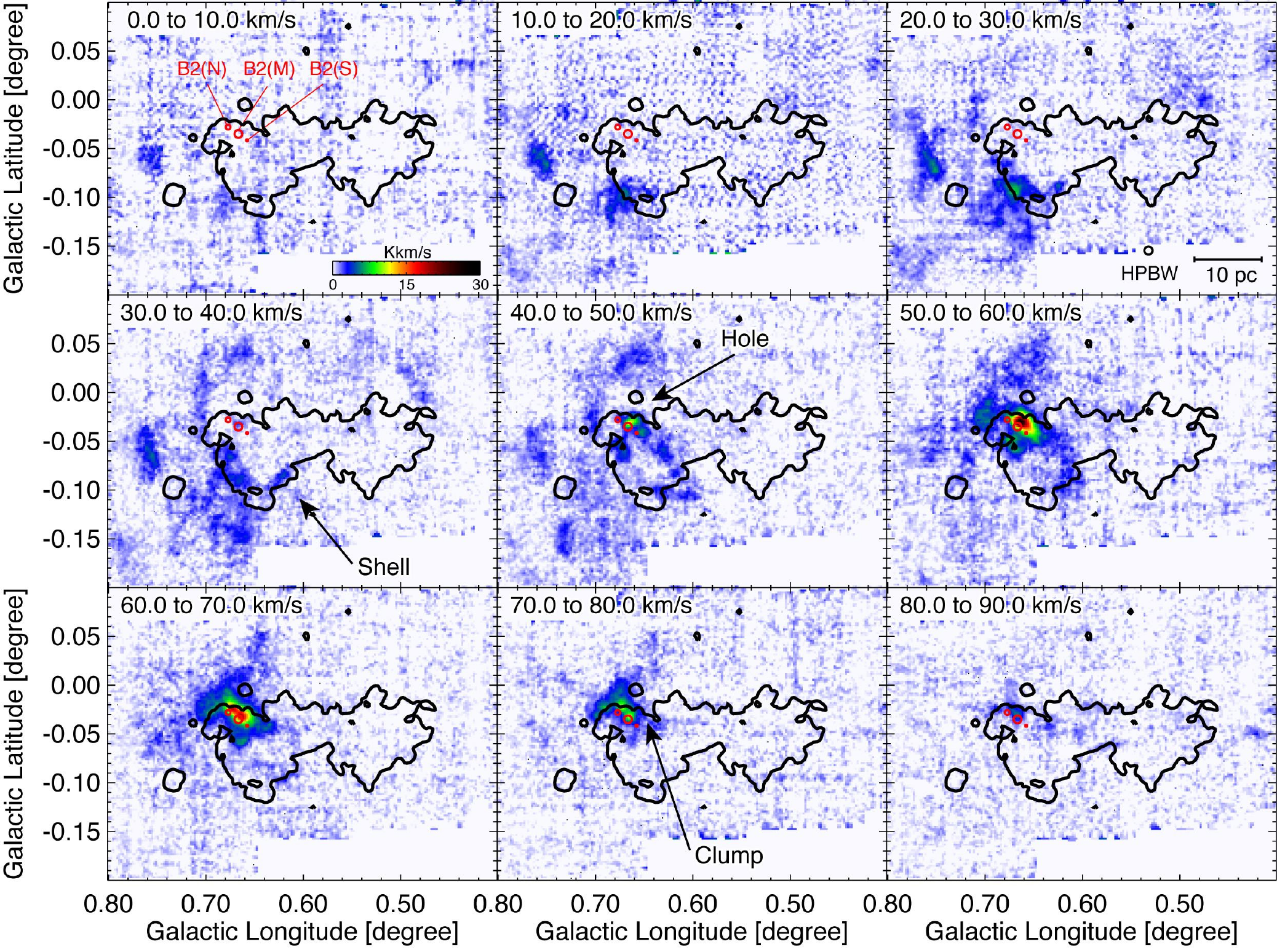}
  \caption{Velocity channel distributions of SiO($J$=5--4) toward the Sgr\,B region, obtained with APEX, overlaid with the thick black contours of the 90 cm radio-continuum emission. The red circles indicate the positions of Sgr\,B2(N), (M), and (S).}
  \label{sio}
  \end{center}
\end{figure*}

\begin{figure*}[t]
\begin{center}
    \includegraphics[width=16cm]{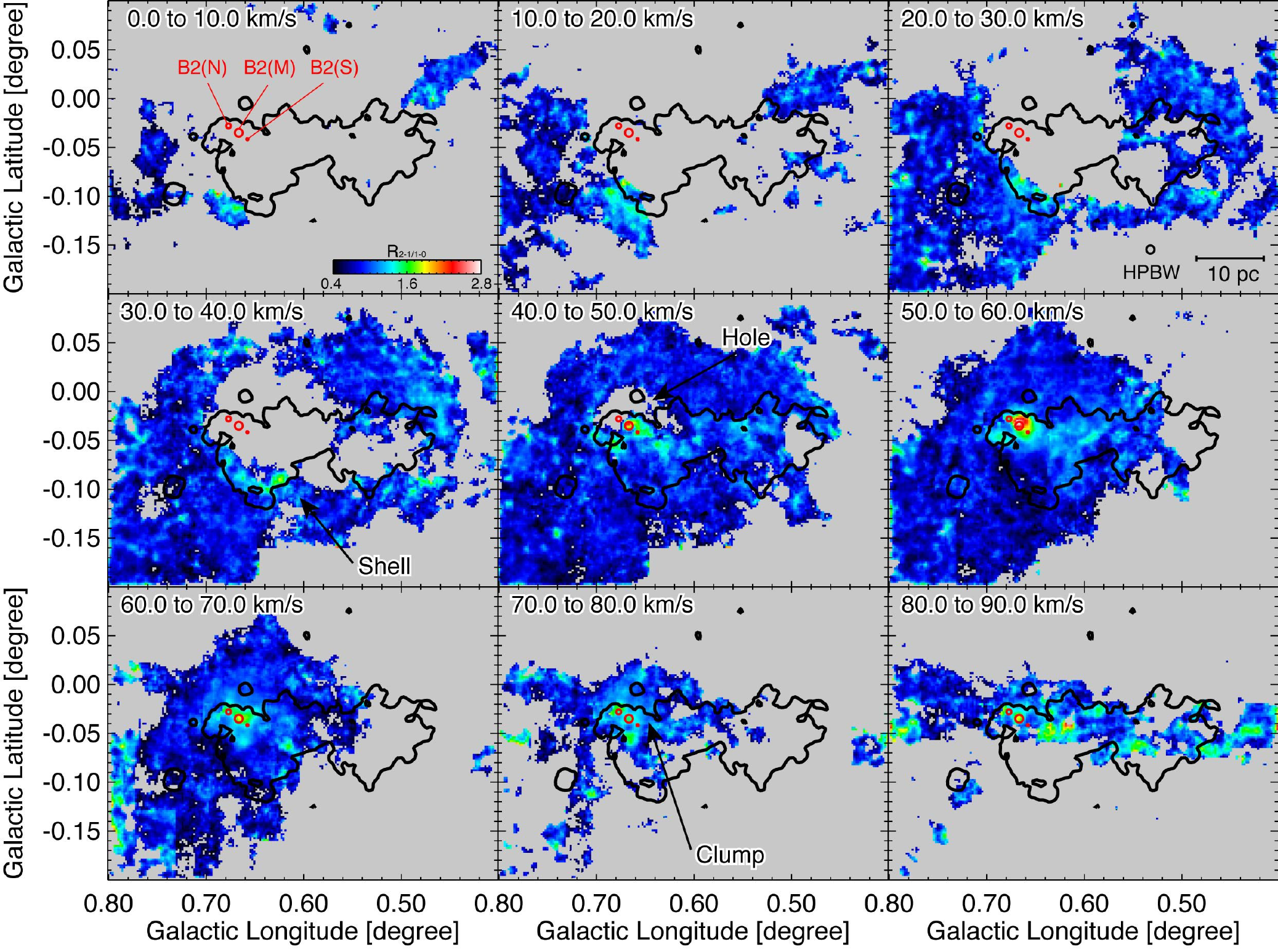}
  \caption{Velocity channel distributions of the intensity ratios of $\ceighteenoh$ and $\ceighteenol$ toward the Sgr\,B region, overlaid with the thick black contours of the 90 cm radio-continuum emission. The red circles indicate the positions of Sgr\,B2(N), (M), and (S).}
  \label{ratio}
  \end{center}
\end{figure*}

Here we show the distributions of the shocked gas (Figure~\ref{sio}) and the relative excitation states (Figure~\ref{ratio}) toward the Sgr\,B1 and Sgr\,B2 regions.
Figures~\ref{sio} and \ref{ratio} show the velocity channel distributions of $\sioh$ and $\rco$, respectively.

\bibliography{ref_sgrb_enokiya}{}
\bibliographystyle{aasjournal}

\end{document}